\documentclass[prb,aps,superscriptaddress,preprint,showpacs]{revtex4-1}

\usepackage{graphicx}
\usepackage{amsmath}
\usepackage{amssymb}
\usepackage{gensymb}

\usepackage{hyperref}

\begin{document}

\title{Dipolar ferromagnetism in three-dimensional superlattices of nanoparticles}

\author{Bassel Alkadour}
\affiliation{Department of Physics and Physical Oceanography, Memorial University, St. John's, NL, Canada A1B 3X7} 
\affiliation{Department of Physics and Astronomy, University of Manitoba, Winnipeg, MB, Canada R3T 2N2}

\author{J. I. Mercer}
\affiliation{Department of Computer Science, Memorial University, St.~John's, NL, Canada, A1B 3X5}

\author{J. P. Whitehead}
\affiliation{Department of Physics and Physical Oceanography, Memorial University, St. John's, NL, Canada A1B 3X7} 

\author{B. W. Southern}
\affiliation{Department of Physics and Astronomy, University of Manitoba, Winnipeg, MB, Canada R3T 2N2}

\author{ J. van Lierop}
\affiliation{Department of Physics and Astronomy, University of Manitoba, Winnipeg, MB, Canada R3T 2N2}

\date{\today}

\begin{abstract}
A series of atomistic finite temperature simulations on a model of an FCC lattice of maghemite nanoparticles using the stochastic Landau-Lifshitz-Gilbert (sLLG) equation are presented. The model exhibits a ferromagnetic transition that is in good agreement with theoretical expectations. The simulations also reveal an orientational disorder in the orientational order parameter for $T < 0.5 T_c$  due to pinning of the surface domain walls of the nanoparticles by surface vacancies. The extent of the competition between surface pinning and dipolar interactions provides support for the conjecture that recent measurements on systems of FCC superlattices of iron-oxide nanoparticles provide evidence for dipolar ferromagnetism is discussed. 
\end{abstract}

\maketitle

\section{Introduction}

Magnetism has provided a fertile field in understanding emergent behaviour resulting from strong correlation effects via the signatures of symmetry breaking. While much of the current research in this area is focused on exchange-type driven interactions that arise from the strong correlation effects amongst electrons (e.g. topological insulators), the dipolar interaction, the poor cousin of exchange, also provides examples of novel forms of emergent behaviour (e.g. stripe phases\cite{stripe} and spin ices\cite{ice}). 

From a theoretical perspective, the dipolar interaction is especially appealing as it involves no ``adjustable parameters", it has no intrinsic length scale and, depending on the context, can be either ferromagnetic or antiferromagnetic. In the seminal works of Luttinger and Tisza\cite{Luttinger.1946} and Kittel\cite{Kittel.1951} it was shown that while the dipolar interaction can, in and of itself, give rise to magnetic order in a three dimensional lattice of point dipoles, it is ferromagnetic only in FCC and BCC structures (with magnetic dipole domains and domain walls) and antiferromagnetic in all others. Despite the obvious significance of magnetic order emerging from dipolar interactions, the experimental verification of this result has been elusive as there is a paucity of real systems in which the dipolar interaction dominates exchange interactions at the atomic or molecular level. 

A notable exception to this is the arrays of synthetically produced single domain magnetic nanoparticles.  While there is now a substantial body of work on well-defined magnetic nanoparticles and their intrinsic nanomagnetism, and it has been shown definitively that their large (effective) dipole moments result in strong interparticle interactions yielding novel physics in disordered systems at high densities\cite{Yamamoto.2008, Kasyutich.2010, Yang.2014}, experiments on superlattice crystals of nanoparticles (e.g. in FCC arrangements over micron length scales\cite{Kasyutich.2010, johan3d, Yang.2014}) have revealed what can be considered as only tantalizing hints of dipolar ferromagnetism -- the collective, cooperative behaviour is still poorly understood. This is due in large part to the subtle interplay between intra-(atomic spin) and interparticle magnetism\cite{Kasyutich.2010, Yang.2014, Desvaux.2005, DeToro.2013, Faure.2013}. 

In this paper we present results from a series of atomistic finite temperature simulations on a model of an FCC lattice of maghemite nanoparticles using the stochastic Landau-Lifshitz-Gilbert (sLLG) equation. These simulations, when taken together with those presented in our earlier studies on ensembles of non-interacting maghemite  nanoparticles\cite{rapidComm2016}, provide an interesting complement to recent experiments on magnetoferritin (Fe$_3$O$_4$/$\gamma$-Fe$_2$O$_3$) particles\cite{johan3d} in which the particles can be self-assembled to form a FCC superlattice with typical length of the order of 1.5-2.0 $\mu$m but which can be disassembled following the application of an optical stimulus. Comparing the results from before and after the disassembly clearly show the effects of the FCC ordering on the magnetic properties of the particles. The extent to which these simulation studies support the conjecture that the observed differences serve as a signature of dipolar ferromagnetism, and how further study might help resolve this question, is discussed.

\section{FCC point dipole lattice}

Theoretical studies of ordered arrays of point dipoles  with uniform magnetizations interacting only through dipole interactions have a long history, with particular interest surrounding the prediction of long-range ferromagnetic order in  the case of particles in an FCC lattice configuration\cite{gse, fcc-monte}. The energy of a point dipole lattice can be written as
\begin{equation}
E_d= g  \sideset{}{'}\sum_{\langle ij\rangle} 
  \left( \frac{{\hat{\sigma}}_{i}\cdot {\hat{\sigma}}_{j}}{r_{i j}^3} -3  \frac{({\hat{\sigma}}_{i}\cdot \vec{r}_{i j})
({\hat{\sigma}}_{j}\cdot\vec{r}_{i j})} {r_{i j}^5}\right)
\label{eq:dipoleEnergy}
\end{equation}
where $\hat{\sigma}_i$ are unit vectors defined at each site $i$, $\vec r_{ij}=\vec R_{ij}/a$ is the relative position of two dipoles in units of the nearest neighbour separation $a$ and the sum is over all pairs of atoms $\langle ij \rangle$  with $i\ne j$. The coupling $g=\mu_0 m^{2}/4\pi a^{3} k_B$~Kelvin, where $m$ denotes the magnitude of the dipole moment on each site. 

As part of this study, simulations on a FCC lattice of point dipoles with periodic boundary conditions were also performed using sLLG and finite size scaling applied to various thermodynamic quantities.  A comparison between our later simulation results for the FCC nanoparticle array and this equivalent point dipole FCC lattice will provide corroboration of the simulation results for the nanoparticle array magnetism as we would expect them to agree, qualitatively at least, to leading order. It will also serve to distinguish those properties that may be attributed to the subtle interplay between the internal magnetic structure of the nanoparticles and the FCC lattice internal dipole field.

\begin{figure}[b!]
\begin{center}
\includegraphics[scale=0.7]{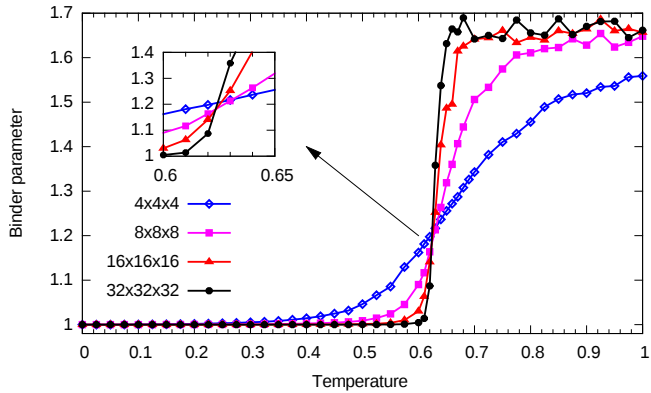}
\caption{(Color online) Binder ratio ${\langle M^4 \rangle}/{\langle M^2 \rangle^2}$ for different sizes of the FCC lattice of magnetic dipoles as a function of reduced temperature.\label{fig:FCCb}}
\end{center}
\end{figure}

Of particular relevance to the current work is the accurate determination of the Curie temperature $T_c$ of an FCC  lattice 
of point dipoles. Fig.~\ref{fig:FCCb} shows the Binder ratio $\langle M^4 \rangle / \langle M^2 \rangle^2$ of the calculated 
overall magnetization for linear sizes $L=4, 8, 16$ and $32$.  We find a ground state energy of $E_0/g= -2.962= -2\pi \sqrt{2}/3$ and a transition temperature of $T_c=T/g=0.625$ with the magnetization oriented along the [111] axis, in good agreement with 
previous results\cite{gse,matthew,fcc-monte,gringas2014}.  

It is well established that the  ground state energy of a saturated classical face-centered cubic dipole lattice is independent of the 
orientation of the magnetization. This degeneracy can be lifted by fluctuations at finite temperature through the mechanism of order by disorder\cite{villain,henley} and the magnetization axis is determined by an effective induced anisotropy. However, while previous simulation studies by Bouchaud and Zerah\cite{fcc-monte} reported the existence of a reorientation transition 
from the $[111]$ axis to the $[100]$ axis at approximately $T\sim T_c/2$, we found no indication of such a transition in any of 
our simulations. The absence of such a transition has been confirmed by other independent simulation studies\cite{gringas2014}.


\section{Multiscale model}

The simulations for the nanoparticle superlattice were performed on a model consisting of 512 ($8\times 8\times 8$) spherical maghemite ($\gamma$-Fe$_2$O$_3$) nanoparticles on an FCC lattice with periodic boundary conditions. The nanoparticles are represented by an atomistic core-shell model consisting of approximately 8200 Fe$^{3+}$ spins in which the core has bulk-like exchange and the shell has weak exchange and radial anisotropy as described in Ref.~\citenum{rapidComm2016}. The model also includes the dipolar interactions between the nanoparticles calculated self consistently using a multi-scale approach that is naturally parallelizable. This represents a more fundamental approach than the more phenomenological models composed of a system of coupled superspins\cite{modelMeiklejohnBean, plumer2010, brinis2014, cubicAniso}. The calculations were all performed using the micromagnetics scripting language MagLua\cite{basselqd2016} that has been successfully applied to a number of atomistic and micromagnetic simulation studies in nanomagnetism\cite{HAMR2011, spinWavesLLG2011, kmcThermalDecay2013}.

Results are presented for nanoparticles of diameter $D$=$7.5$~nm with core diameters $D_c$=6.3~nm and 6.75~nm, which we refer to as the FCd63 and FCd675 superlattices respectively. The nanoparticles are single crystals with a total of 382 spinel unit cells. The numbers of core cells are 226 and 278 and the number of surface cells are 156 and 104  for FCd63 and FCd675 respectively. All of the surface cells are incomplete since they are cut by the particle radius whereas the core cells are all complete. Both the FCd63 and FCd675 superlattices show an order/disorder transition from a superparamagnetic configuration to a true ferromagnetic state (i.e. not superferromagnetism\cite{Venero.2016, Bedanta.2007} where exchange interactions amongst nanocrystallites dominate instead of the much weaker dipolar interactions) at a temperature that is consistent with theoretical expectations\cite{fcc-monte}. 

To simulate a model comprising $N=512$ (8$\times$8$\times$8 lattice) nanoparticles each consisting of approximately 8200 Fe$^{3+}$ ions at an atomistic level that includes the full dipolar interaction using sLLG is, currently, simply not feasible.  Instead, we employ a multiscale model in which we assume that the magnetic intraparticle interactions are dominated by exchange and a single-site anisotropy while the interparticle interactions consist solely of the dipolar interaction. The  dipolar interaction energy, Eqn.~\ref{eq:dipoleEnergy}, in this multiscale approximation then simplifies to
\begin{equation}
 E_\mathrm{eff} =g \sideset{}{'}\sum_{\langle kl\rangle}  \left( \sum_{i\in k} \hat{\sigma}_{ki}\right) \cdot\mathbf{\Gamma}^{kl} 
 \cdot \left(\sum_{j\in l} \hat{\sigma}_{lj}\right)
\label{eq:nshd}
\end{equation}
where the subscripts $\{ki\}$ denote the $i^\mathrm{th}$ spin in the $k^\mathrm{th}$ nanoparticle and $\mathbf{\Gamma}^{kl}$ is the interaction tensor between point dipoles located at the centre of each nanoparticle located on a FCC lattice and calculated assuming periodic boundary conditions using the Ewald summation technique. This multiscale approach considerably reduces the computational effort required while retaining the complexity of the spin structure of the individual nanoparticles in combination with the long range dipolar interactions between the nanoparticles.

Each of the nanoparticle's Fe$^{3+}$ atoms has a moment of $m$= 5$\mu_B$. The spinel lattice structure of the $\gamma$-Fe$_2$O$_3$ nanoparticles has tetrahedral (A) and octahedral (B) sites which order ferrimagnetically with a net moment of $1.25 \mu_B$ per atom in the bulk. In order to maintain charge neutrality, 1/6 of the octahedral sites are occupied by vacancies\footnote{While in the simulation results reported in this study, the spinel structures of the individual nanoparticles were aligned along a common axis, additional simulations runs in which the orientations were randomly assigned showed no detectable differences for the systems of interest in the present work.}. It is assumed that the surface spins, defined as those spins located in the region $D_c/2 < r < D/2$, experience a radial single site anisotropy due to the broken translational symmetry. The proportion of surface spins is around $41\%$ and $27\%$ for the FCd63 and the FCd675 superlattices, respectively. The exchange coefficients are those used in Ref.~\citenum{rapidComm2016} with the surface anisotropy constant $K_s/k_B = 10\,\mathrm{K}$.  Periodic boundary conditions were applied to the 8$\times$8$\times$8 array. 

The simulations were parallelized such each of the individual nanoparticles on the FCC array was assigned to a single processor core. The dipole fields at each site on the FCC lattice were calculated and communicated using the Message Passing Interface protocol (MPI).  The sLLG time steps were $\Delta t$ = $2 \times 10^{-4}$~$t_u$, where the time unit $t_u$=1~Tesla/$\gamma$=5.68$\times 10^{-12}$~sec, and the damping factor $\alpha = 0.5$. Since the dipole field changes were very small in a single sLLG time step, tests showed that the dipole field only needed to be updated every 100$\Delta t$ with no measurable loss in accuracy. 
  
We represent the $i^\mathrm{th}$ spin in the $k^\mathrm{th}$nanoparticle by a unit spin $\hat{\sigma}_{ki}$. We calculate the average magnitude of the normalized magnetization associated with the individual nanoparticles ($M_n$) and the average magnitude of the normalized magnetization of the entire nanoparticle lattice ($M_{nl}$) using the definitions
\begin{eqnarray}
M_{n}(T) \!&=&\! \frac{4}{N}  \sum\limits_{k=1}^{N} \left| \sum\limits_{i=1}^{q_{k}} \frac{ \hat{\sigma}_{ki}} {q_{k} } \right|~~\\
M_{nl}(T) \!&=&\! \frac{4}{N}  \left| \sum\limits_{k=1}^{N} \sum\limits_{i=1}^{q_{k}} \frac{ \hat{\sigma}_{ki}} {q_{k}} \right| 
\end{eqnarray}
where $N$ is the number of nanoparticles in the lattice and $q_k$ is the number of spins in the $k^{th}$ nanoparticle. 
These moments are normalized to the bulk ferrimagnetic magnetization of each nanoparticle, and are equal to one when 
the $\{\hat\sigma_{ki}\}$ are aligned along a common axis with the A-site spins anti-parallel to the B-site spins. 
The factor of $4$ is needed to normalize the moments since, as mentioned above, the maximum magnetic moment per site for a ferrimagnetic nanoparticle is $1/4$ that of the moment on each Fe atom.
 
We also define the equivalent point dipole FCC lattice consisting of $N$ sites at which is located a temperature dependent dipole moment $\vec m_k = m_n(T) \hat \sigma_k$ ($ 1 \le k \le N$) where $m_n(T)=1.25 \mu_B M_n(T)\langle q_k \rangle$ is the average magnetic moment of the nanoparticles and $\hat{\sigma}_k$ is a unit vector defining the orientation of the dipole moment $\vec m_k$.  The energy of the equivalent dipole lattice is therefore given by  Eqn.~\ref{eq:dipoleEnergy} with $g\to \tilde g(T) = \mu_0 m_n^2(T)/4\pi a^3 k_B$. The magnitude of the normalized magnetization of the equivalent dipole lattice is defined as
\begin{equation} 
M_{dl}=(M_n(T)/N) |\sum_{k=1}^{N} \hat{\sigma}_k |.
\end{equation}


In what follows we have assumed that the ratio of the nanoparticle diameter to the FCC lattice spacing $r = D/a$ is equal to unity. This is an idealization of the experimental situation\cite{Kasyutich.2010} which consists of synthesized magnetoferritin nanoparticles with a iron-oxide diameter of $\sim$7~nm encapsulated in a 1~nm protein layer; hence $a = 8.5$ nm. Our choice of $D = 7.5$ nm  with $r=1$ captures the interplay between the long range character of the dipolar interaction between the nanoparticles and their internal spin structure that is the focus of the current work. The presence of the magnetoferritin coating however is important in that it justifies the absence of any effective exchange coupling between the nanoparticles in our model.

In order to provide some insight into the dipole ordering of a single nanoparticle, consider the situation where maghemite has no exchange and is a pure dipolar system. The parameter $g$ in Eqn.~\ref{eq:dipoleEnergy}
would have the value $g\sim 0.08$ Kelvin. For an FCC lattice structure this would give a dipole ordering temperature of $T_c\sim 0.05$ Kelvin. By contrast,  the large net moment of the maghemite NPs yields a value of $\tilde g \approx 83\,  \mathrm{K}$ and $58\,  \mathrm{K}$ for the FCd675 and FCd63 systems respectively\cite{bassel2016}. This corresponds to $T_c \approx 52\,  \mathrm{K}$ (FCd675) and  $36.5\,  \mathrm{K}$ (FCd63). 

\section{Multiscale simulation results}

\begin{figure}[b!]

\begin{center}
\includegraphics[scale=0.65]{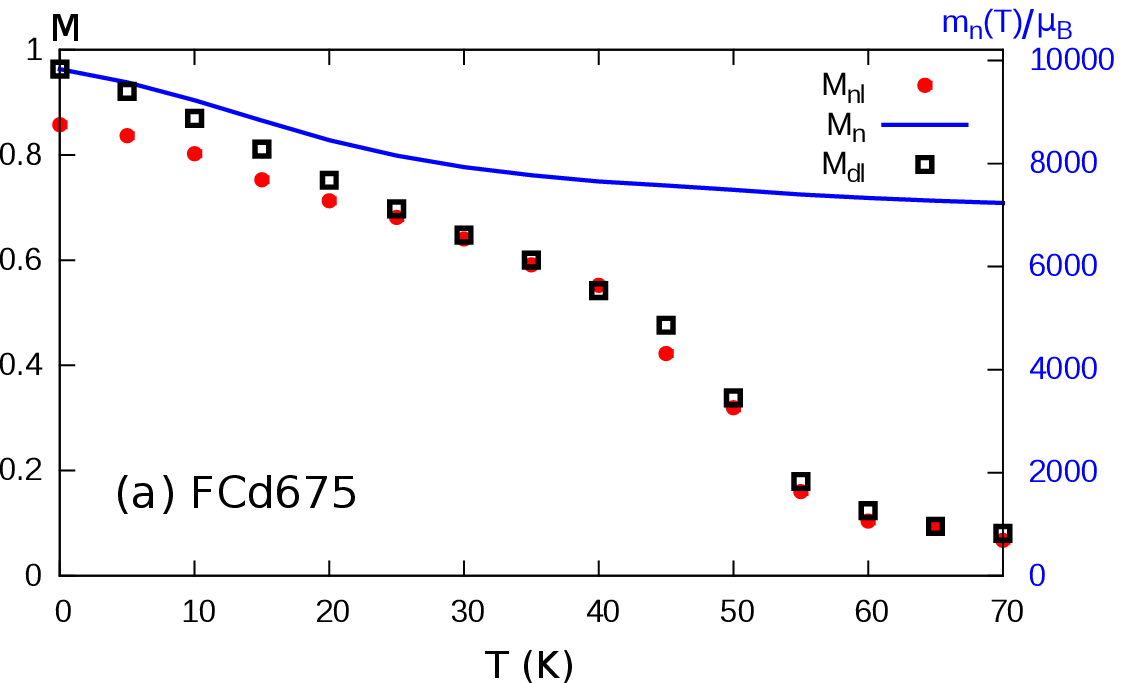}

\includegraphics[scale=0.65]{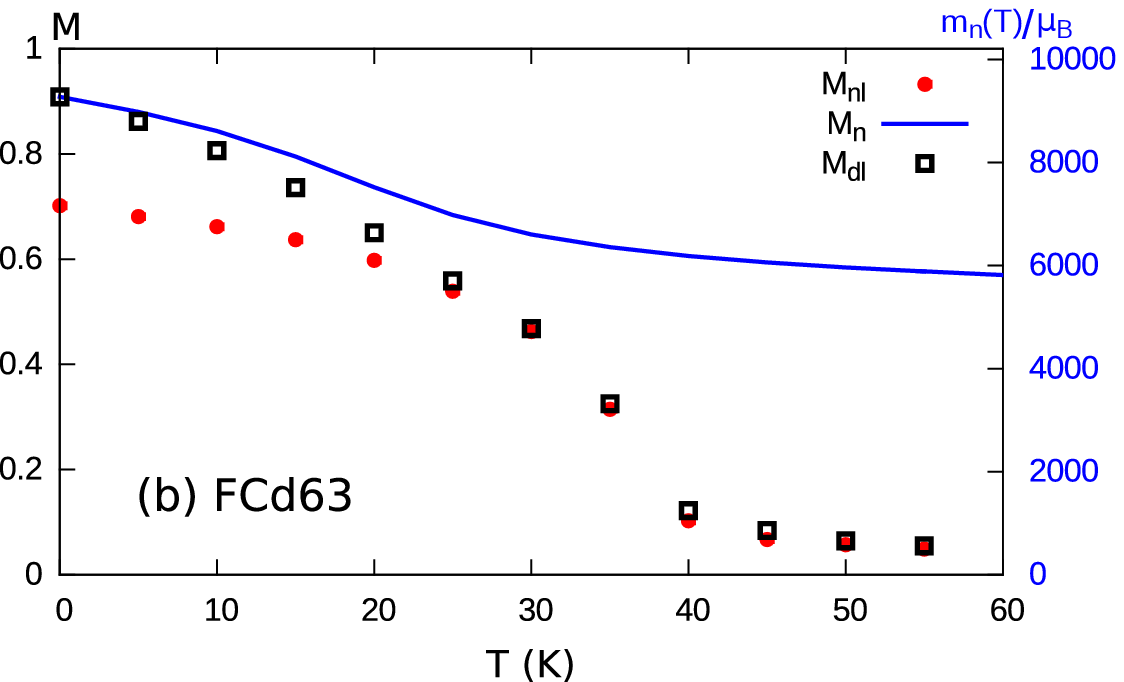}
\end{center}

\caption{(Color online) The normalized magnitudes of $M_{nl}$, $M_{n}$, $M_{dl}$ as defined in Eqns.(3), (4) and (5) are presented as a function of temperature. The right axis indicates the magnitude of the average dipole moment of the nanoparticles $m_n/\mu_B$ in units of Bohr magnetons. The upper panel (a) is for the FCd675 system and the lower panel (b) for the FCd63 system. Note the different temperature scales between a) and b)\label{fig:M-675}}
 \end{figure}

Our simulation results are presented in Fig.~\ref{fig:M-675} for both arrays. The average magnitude of the nanoparticles' magnetization, $M_{n}$, exhibits a smooth variation with temperature up to approximately 900~K (not shown), above which it is effectively zero. The increase in $|dM_n(T)/dT|$ for both samples below $\sim$30~K is due to the partial ordering of the surface moments\cite{rapidComm2016}. The frustration of the surface spins due to the radial anisotropy prevents the complete saturation of the nanoparticles' magnetization. The open squares indicate the magnetization, $M_{dl}$, of  the equivalent dipole lattice for which each site has a moment of magnitude $M_n$ at each temperature. The solid dots represent the magnitude of the lattice magnetization, $M_{nl}$, obtained using our multiscale approach. The left axis indicates the normalized values of the moments and the right axis indicates the values of the moments in $\mu_{\rm{B}}$. The dipole interactions included in the nanoparticle arrays show negligible effect on $M_{n}$.

\begin{figure}[t!]

\begin{center}
\includegraphics[scale=0.7]{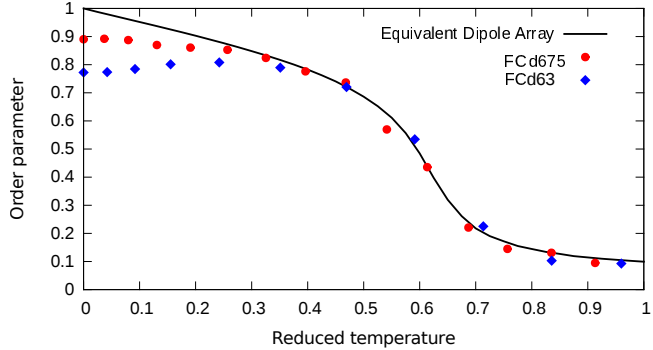}
\end{center}

\caption{(Color online) The order parameter  $\sigma_{nl}$  for the FCd63  (blue diamonds) and FCd675 (red circles)  plotted for 8$\times$8$\times$8 FCC arrays as a function of reduced temperature $\tilde T$. Also shown is the order parameter for the equivalent dipole lattice (solid line) as a function of its reduced temperature $T/g$.\label{fig:O-rk4}}
\end{figure}

The data show that the nanoparticle superlattices and the equivalent point dipole system begin to order ferromagnetically along the $[111]$ axis at $T_c$$\approx$55~K and 40~K for the FCd675 and the FCd63 superlattices, respectively.   The magnetization for both the FCd675 and the FCd63 superlattices, ($M_{nl}$) and the equivalent dipole lattice ($M_{dl}$) track each other until $\approx$20~K, below which the surface spins start to order and $M_{nl}$ drops below $M_{dl}$. The increased disorder at low temperatures is more obvious if we eliminate the effect of the temperature dependence of the magnitude of the magnetization of the nanoparticles by plotting the order parameter defined by   $\sigma_{nl}=M_{nl}/M_{n}$ as a function of the reduced temperature $\tilde T =  T/\tilde g(T)$, as shown in Fig.~\ref{fig:O-rk4}. Note that in the case of the FCd63 superlattice (smaller core nanoparticles) the order parameter, $\sigma_{nl}$, actually decreases with decreasing temperature for $\tilde T < 0.4$. This discrepancy between the equivalent dipole lattice and the nanoparticle superlattices indicates that there is some phenomenon that decreases the orientational order between the nanoparticles.

The origin of this additional disordering implied by the reduced order parameter $\sigma_{nl}$ observed in Fig.~\ref{fig:O-rk4} may be understood  from our  previous simulation studies on ensembles of non-interacting maghemite nanoparticles\cite{rapidComm2016}, in which it was shown that the magnetic moment of the nanoparticles were pinned by the surface vacancies in the octahedral B-sites by the surface magnetization. This pinning effect is a result of the frustration that arises as consequence of the competition between the exchange and the surface anisotropy. Because of this competition it was shown in Ref.~[\onlinecite{rapidComm2016}] that for each nanoparticle there exists a N\'{e}el-like domain wall in the surface magnetization at the equatorial plane separating the north and south magnetic poles in which the spins, located at the sites within this domain wall, are highly frustrated. As a consequence the presence of a vacancy located at a site within the domain wall will generally result in a lower energy than if the same vacancy were located at a site close to one of the poles. 

\begin{figure}[htp!]
\centering 
\includegraphics[width=0.5\textwidth]{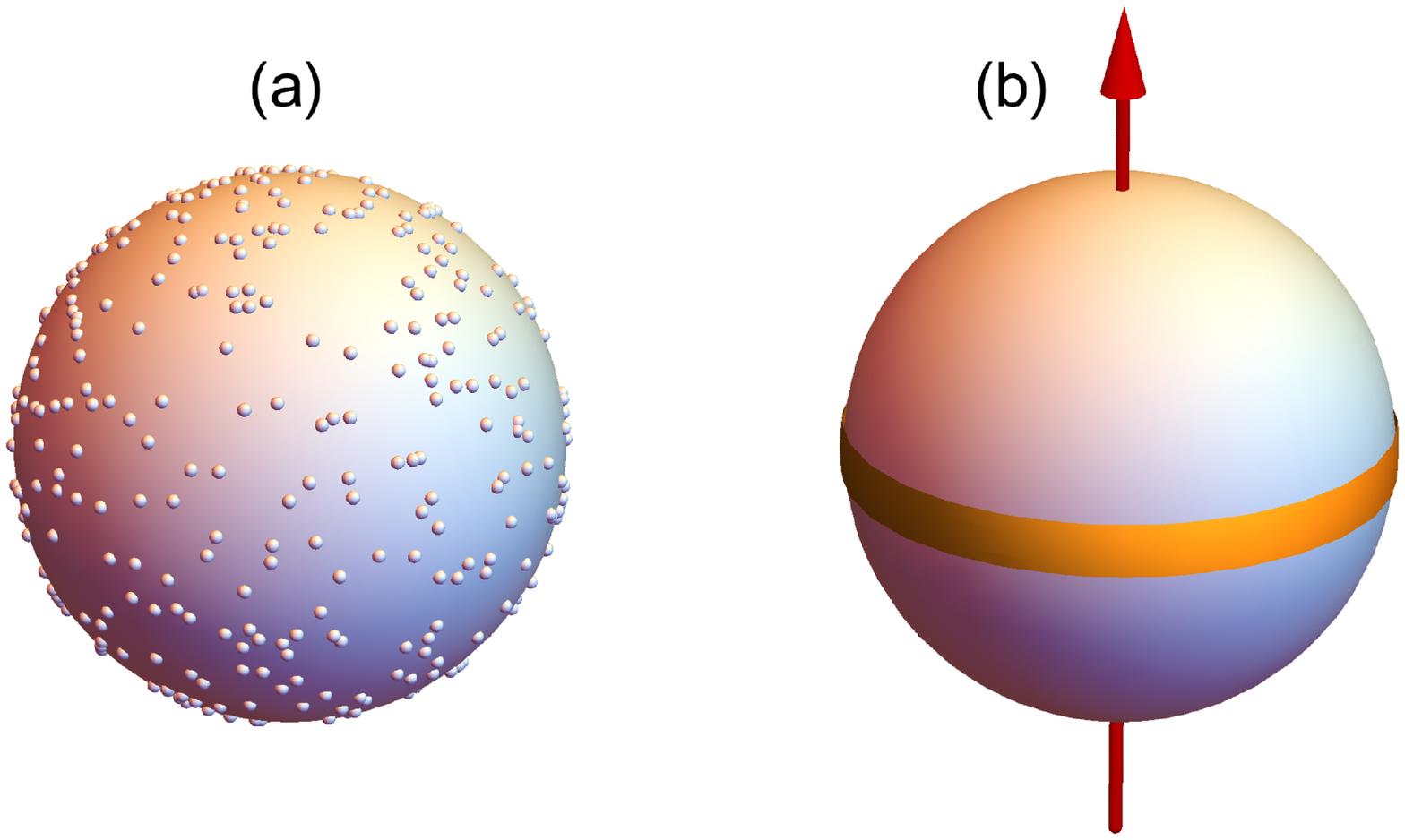}
\includegraphics[width=0.5\textwidth, trim=0mm 0mm 22.5mm -10mm]{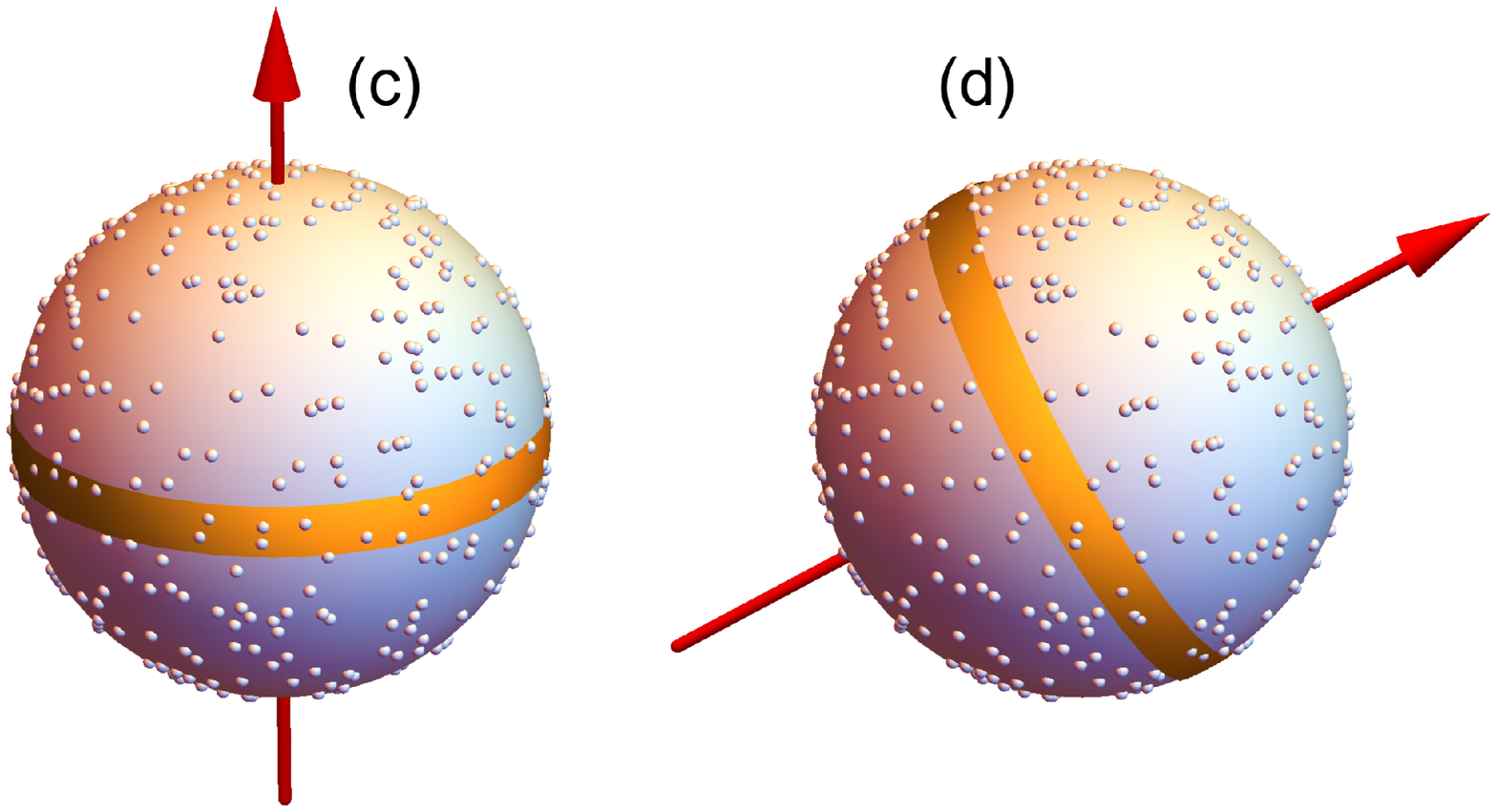}
\caption{\label{fig:simulatedPinning} (a) A figure showing 500 points distributed randomly over the surface of the unit sphere. The points represent the location of the surface vacancies in the octahedral B-sites. (b) A schematic plot illustrating the direction of the NP magnetic moment, in this case aligned along the $z$-axis, with the shaded region representing the domain wall located at the equator separating the north and south magnetic poles. (c)  Figures (a) and (b) superposed to illustrate schematically the surface vacancies located within the domain wall for the case of the NP moment aligned along the $z$-axis. (d) A schematic plot similar to (c) but with the magnetic moment and the domain wall rotated by 60\degree\;  about the $x$-axis. }																	
\end{figure}

To understand the role of these surface vacancies it is important to keep in mind that, as indicated in Ref. [\onlinecite{rapidComm2016}], while the distribution of the vacancies among the B-sites is statistically uniform, statistical variance and the crystallographic structure of the nanoparticle will give rise to a spatial clustering of the vacancies. Because of this clustering the number of surface vacancies contained within the domain wall will depend on the orientation of the equatorial plane. While the locations of the vacancies are fixed the location of the equatorial plane, oriented perpendicular to the magnetic moment, is not and hence the number of vacancies located within the domain wall will depend on the direction of the magnetic moment of the nanoparticle. This is illustrated schematically in Fig.~\ref{fig:simulatedPinning}  in which 500 randomly distributed points located on the surface of the unit sphere representing the surface vacancies of a magnetic NP are plotted (Fig.~\ref{fig:simulatedPinning}(a)). The points were generated using the Marsaglia algorithm\cite{marsaglia1972,muller1959} . Also shown is a schematic representation of the magnetic moment vector of a magnetic nanoparticle aligned along the $z$-axis (Fig.~\ref{fig:simulatedPinning}(b)) with the shaded region, defined by $-w/2 < z <w/2$ with $w= 0.2$,  representing that portion of the NP surface occupied by the domain wall. In Fig.~ \ref{fig:simulatedPinning}(c) the vacancy distribution in (a) is shown superposed on the magnetic moment in (b). The number of surface vacancies contained within the domain wall for this particular distribution is calculated to be $N_v=43$. In Fig.~\ref{fig:simulatedPinning}(d) we show the same distribution of surface vacancies as in (b) but in this case superposed on schematic representation for the magnetic moment vector of the NP rotated by 60\degree\; about the $x$-axis. Counting the number of vacancies within in the rotated domain wall we obtain $N_v = 33$. While the surface vacancy distribution in this illustrative example is much simpler than that of the model NP used in our simulations, it nevertheless serves to highlight two key points. Firstly, the fact that, even although the distribution of surface vacancies is statistically uniform, the effects of clustering due to statistical invariance can nevertheless be significant. The effects of the crystallographic structure of the NP and the finite thickness of the surface, ignored in this simple model, will only serve to enhance the clustering. Secondly, the model explicitly demonstrates the fact that while the locations of the vacancies are fixed, the polar distribution of the vacancies, measured with respect to the axis aligned parallel to the NP magnetic moment vector, will depend on the orientation of the NP magnetic moment. The combination of the arguments that (a) the number of vacancies contained within the vicinity of the magnetic equator depends on the orientation of the NP magnetic moment and (b) that surface vacancies located in the vicinity of the equator will, due to the effects of frustration, result in a lower energy than those located close to one of the poles gives rise to a non-trivial dependence of the nanoparticle energy on the orientation. These arguments lead the conclusion, first presented in Ref.~[\onlinecite{rapidComm2016}], that the nanoparticle energy will have a minimum when the magnetic moment is oriented so that the number of vacancies in the domain wall region is a maximum, a mechanism that corresponds to the pinning of the nanoparticle magnetic moment, (or perhaps more precisely the pinning of the surface domain wall) by the surface vacancies. The precise dependence of this energy on the orientation of the magnetic moment of the nanoparticle will depend not only on the spatial distribution of the surface vacancies also on the degree of frustration associated with the surface spins located within the domain wall. As such the overall magnitude of the energy variation and the functional form of its relationship to orientation of the magnetic moment will be both temperature and field dependent.

\begin{figure}[t!]
\begin{center}
\includegraphics[scale=0.36]{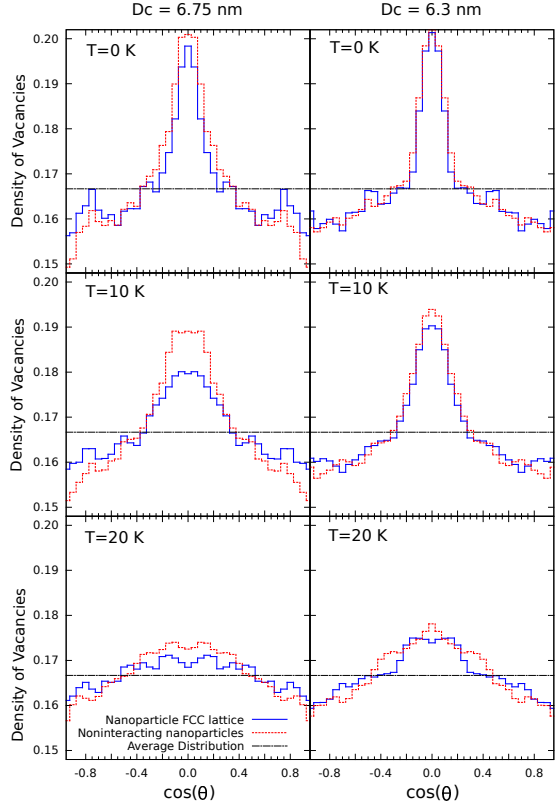}
\end{center}
\caption{(Color online) The average distribution of the surface vacancies for the FCC lattices (solid blue line) FCd675 and FCd63  together with the corresponding distribution for the equivalent ensemble of non-interacting nanoparticles (red dash-dot line) as a function of $\cos({\theta})$ at different temperatures.\label{fig:HVF}}
\end{figure}

Evidence for this pinning in the case of the FCd63 and FCd675 superlattices is shown in Fig.~\ref{fig:HVF} in which  histograms plotting the average number of surface vacancies in the range $z_n\to z_n+\Delta z$ are presented for several temperatures. The data are obtained by averaging the number of surface vacancies in each bin over all the nanoparticles in the ensemble, with the $z$-axis (with $z=\cos\theta$) defined so  that it is aligned parallel to the magnetic moment vector of each individual nanoparticle and passing through its centre as shown schematically (for example) by the arrows in Figs.~\ref{fig:simulatedPinning}(a) and (b)\footnote{The difference in the distribution of vacancies in the north ($z>0$) and south ($z<0$) poles was observed to be statistically insignificant. To reduce the degree of scatter in Fig.~\ref{fig:HVF} the data were also averaged over $\pm z$, which accounts for their apparent symmetry with respect to the $z$-axis.}. For comparison, results for the equivalent non-interacting ensembles as well the for the case of a uniform density are also shown. The peak in the histogram at $z= 0$ indicating an increasing concentration of vacancies at the equator does not, as one might naively suppose, result from motion of either the Fe atoms or the vacancies. The locations of the Fe atoms and therefore the vacancies are assumed to be fixed. Instead it results from changes in the spin configuration of the individual nanoparticles in response to the magnetic forces acting on the individual atoms and the fluctuations induced by the stochastic field, and is a consequence of the highly frustrated nature of the surface spins located in the vicinity of the equatorial domain wall combined with local inhomogeneities within the distribution of the surface vacancies. All the results show an increasing concentration of vacancies at the equator as the temperature falls below 20~K,  at which the surface spins order\cite{rapidComm2016}, a clear signature that the pinning energy increases with increasing surface magnetization. 

The fact that the maximum in the average surface vacancy distributions for both the FCd63 and FCd675 superlattices is slightly lower than for their respective non-interacting ensembles is due to the competition between the dipolar field (absent in the noninteracting ensembles) and the surface vacancies. In addition, the concentration of vacancies at the equator is more pronounced for the FCd63 superlattice than those in the FCd675 superlattice due to the greater fraction of surface spins and the smaller core that enhances the pinning effects of the vacancies at the equator on the overall magnetization alignment, and reduces the magnitude of the nanoparticles' dipole moments.  In addition the smaller core results in a reduced value of $M_n(T)$ and $\tilde g(T)$.  This results in a $T_c$ for the FCd63 superlattice that is  significantly lower than for the FCd675 superlattice, hence much closer to the surface ordering temperature. As a result, the FCd675 superlattice is more ordered than the FCd63 superlattice when the surface spins begin to order and the pinning effect of the surface vacancies activates.

\section{Insights into experiments on FCC superlatticles of nanoparticles}

While the ferromagnetic order observed in these simulations is consistent with theoretical expectations, experimental evidence for such a transition is elusive. Recent experiments\cite{johan3d} directly comparing systems of magnetoferritin (Fe$_3$O$_4$/$\gamma$-Fe$_2$O$_3$) nanoparticles that self assemble to form an FCC superlattice with those obtained after their disassembly following the application of an optical stimuli show significant differences in their magnetic properties that result from the dipolar interaction. However, while it is tempting to assert that these differences may be attributed to the appearance of dipolar ferromagnetism it is by no means obvious that such an assertion is justified. 

In this section we demonstrate that the picture that emerges from these simulations and those presented in Ref.~\citenum{rapidComm2016} are at least qualitatively consistent with the experimental data of Ref.~\citenum{johan3d}, and, more importantly perhaps, how simulations studies similar to those presented in the current work and Ref.~\citenum{rapidComm2016} could be extended to determine more conclusively whether the existence of dipolar ferromagnetism in FCC nanoparticle superlattices can be inferred from existing experimental studies. 

\begin{figure}[b]
\includegraphics[scale=0.5]{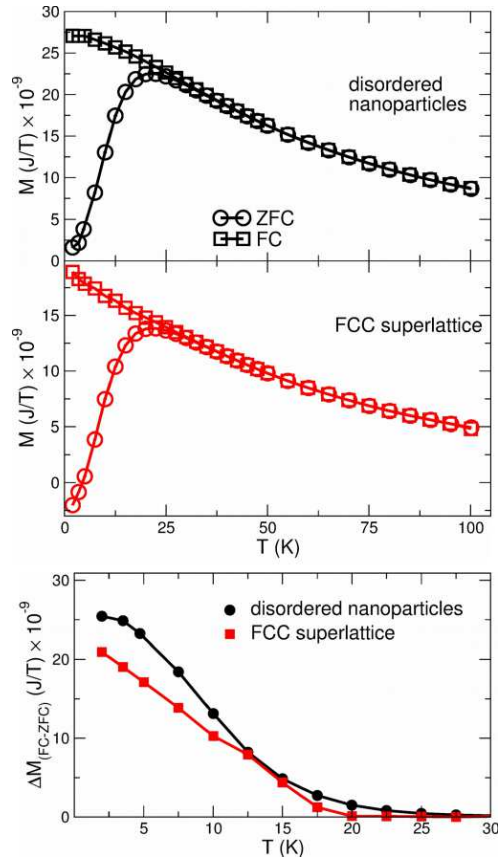}
\caption{(Color online) Top:  Low field ($\mu_oH=10$~mT) magnetization, $M$, in zero field-cooled (ZFC) and field-cooled (FC) configurations for the magnetoferritin nanoparticles in a disordered ensemble and crystals of FCC superlatices.  Bottom:  Difference plot of the zero-field and field-cooled low-field magnetizations, $\Delta M_\mathrm{FC-ZFC}(T)$, for a FCC superlattice of magnetoferritin nanoparticles, and of the same nanoparticles `disassembled' in a disordered ensemble.\label{fig:deltaMvsTdata}\cite{johan3d}}
\label{fig:ZFC}
\end{figure}

Figure \ref{fig:ZFC} shows the heating and cooling of a system of zero field cooled magnetoferritin nanoparticles in the presence of a 10~mT field. These particles can be self-assembled to form a FCC superlattice with typical length of the order of 1.5-2.0~$\mu$m, but which can be disassembled following the application of an optical stimulus, resulting in the typical disordered ensemble of nanoparticles.  Data is shown for both the disordered (disassembled) system and the FCC superlattice. The data for both systems exhibit non-ergodic behaviour over the temperature range 2 to 20~K for the FCC superlattice and 2 to 25~K for the disordered system. In the bottom graph the difference between the magnetization on cooling and heating is also shown. 

The differences in the data clearly show the effects of the dipolar interaction in the case of the FCC superlattice. Of particular interest in the context of the present work is the fact while that the slope $|d\Delta M_\mathrm{FC - ZFC}(T)/dT|$ in the limit  $T\to 0$ tends to zero, in the case of the disordered system, it remains finite in the case of the FCC superlattice. This indicates that while the magnetization is close to saturation in the case of the disordered system, in the case of the FCC superlattice the magnetization is not saturated along the direction of the applied field. 

\begin{figure}
\includegraphics[scale=0.4]{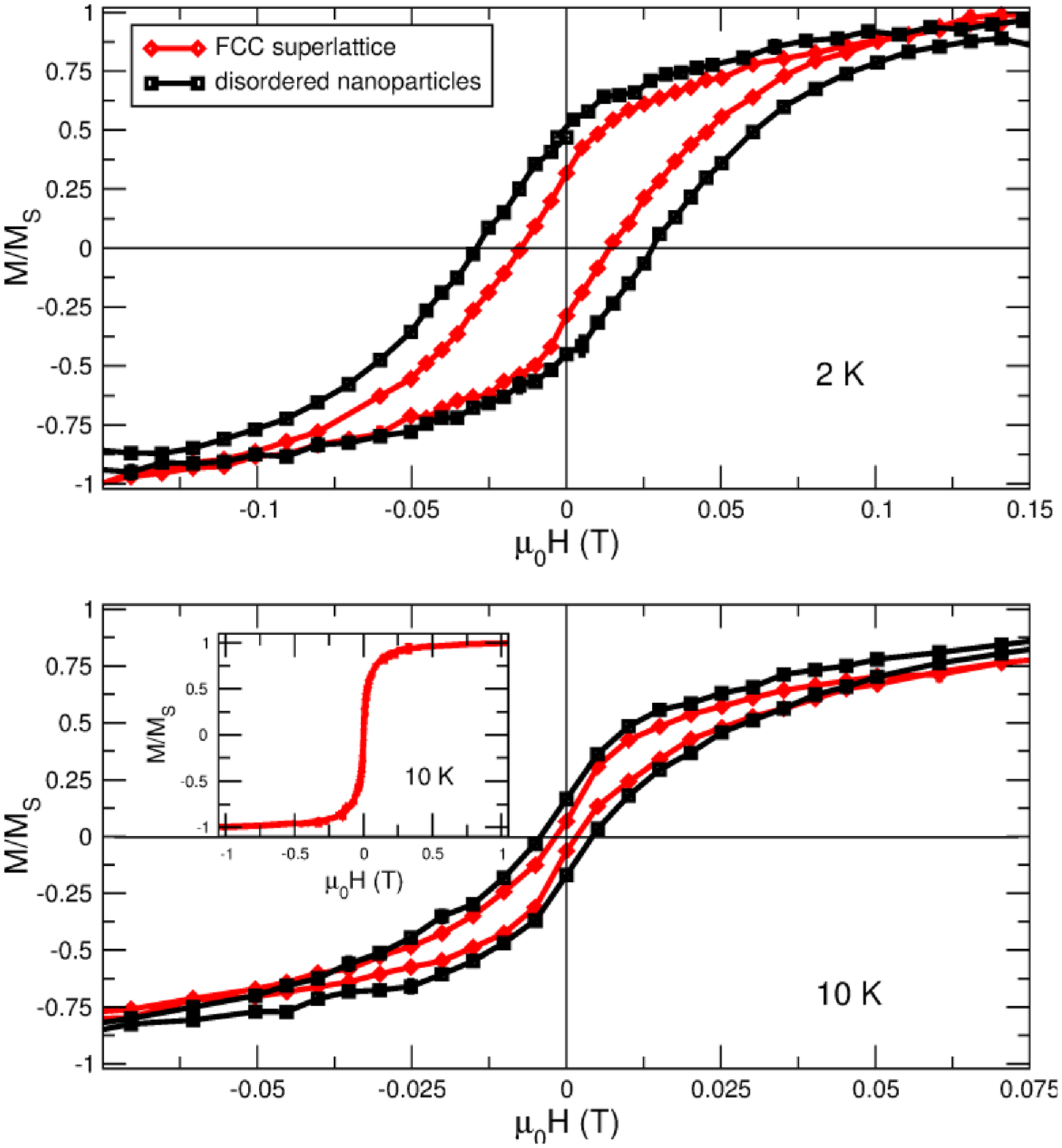}
\caption{(Color online) Field dependent magnetization of 10-25~$\mu$m crystals of the magnetoferritin nanoparticle FCC superlattice ($\diamond$) and the disassembled, disordered magnetoferritin nanoparticles ($\Box$) at 2~K (top) and 10~K (bottom).  The inset show a typical, full $M$ vs  $\mu_0H$ scan of the FCC system.\label{fig:MvsHdata}\cite{johan3d}}
\end{figure}

Figure~\ref{fig:MvsHdata} presents  the normalized magnetization as a function of the applied field at 2 and  10~K for both the FCC superlattice and the disordered system. The principle difference is in the reduced remanent magnetization and the coercivity in the case of the FCC superlattice. This is indicative that the FCC superlattice lattice is composed of system of randomly oriented crystallites.  This feature is consistent with the expectation that in the case the disordered system the moments of the individual nanoparticles will align parallel to the field on cooling and, while the pinning effect will result in some measure of disorder at low temperature, the magnetization will nevertheless be close to saturation in the limit $T\to 0$. In the case of the FCC lattice, on the other hand, the dipole field will dominate and we would expect that, as the system cools, the magnetization would align along the $[111]$ axis that lies closest to the direction of the applied field.  Assuming that the crystallographic axes of the individual magnetoferritin superlattice crystallites are randomly oriented, the net magnetization along the direction of the applied field will not saturate in the limit $T\to 0$. A similar reasoning also provides a plausible explanation for the fact that the normalized remanent magnetization ($M_R = \lim_{H\to 0} M(H)$) obtained from the $M$ vs $\mu_0 H$ loops presented in the upper panel of Fig.~\ref{fig:MvsHdata}  is greater in the case of the disordered system than that observed for FCC superlattice.  Again this may be attributed to the alignment of the magnetization along the randomly oriented $[111]$ axis of the FCC crystallites while in the case of the disordered system the magnetization will be in the direction of the applied field. 

The qualitative nature of the above discussion is an unfortunate consequence of the fact that current computational capabilities limit the time scales that can be accessed by the atomistic theories methods used in this work to the order of ms. This precludes a more a quantitative interpretation of the intrinsically non-equilibrium/non-ergodic behaviour observed in Fig.~\ref{fig:ZFC} and \ref{fig:MvsHdata} based solely on atomistic sLLG simulations. However, as we have shown, such atomistic studies allow us to identify the relaxation processes that dominate at experimental field and temperature sweep rates. In addition, our earlier work\cite{rapidComm2016} presented a simple mean field model of the pinning energy calculated as a function of orientation of the magnetization for a given distribution surface vacancies. Based on this model it is possible to determine the orientation of the metastable spin configurations of a nanoparticle, as well as estimates of the activation energies and attempt frequencies separating them. Such information can provide for a quantitive description of the magnetic properties of magnetoferritin nanoparticles involving experimentally relevant time scales.

\section{Conclusions}

 Our multiscale simulations reveal that, while the nanoparticle superlatttice orders ferromagnetically in accordance with theoretical expectations, it nevertheless exhibits a degree of disorder at low temperatures. It was shown that this disorder is due the pinning effect that arises as a consequence of a subtle interplay between the single-site anisotropy and the vacancies in the region close to the surface, an effect discussed in an earlier paper on ensembles of non-interacting nanoparticles. In addition, we describe how the results from these simulations, combined with those from earlier studies on non-interacting systems of nanoparticles, support the assertion that certain key differences in the magnetic properties of magnetoferritin arrays before and after disassembly from a FCC superlattice may be attributed to dipolar ferromagnetism.  

To provide a more definitive case regarding the experimental verification of emergence of dipolar driven ferromagnetism in FCC nanoparticle superlattices than the above qualitative argument, more detailed simulation studies of non-equilibrium properties  (ie. heating/cooling and $M$ vs $\mu_0 H$ loops) at experimental sweep rates  are required. Such simulation studies, not currently feasible using standard sLLG due to the time scales involved, can play a vital role in this process. We are currently exploring the application of other simulation methods such as Kinetic Monte Carlo\cite{chantrell84,kanai1991,charap1994,fal2013} and Forward Flux Sampling\cite{allen2009,vogler2013} that are potentially applicable to experimentally relevant sweep rates.

\begin{acknowledgments}
This work was supported by the Natural Sciences and Engineering Research Council of Canada,  Compute Canada, ACEnet and  WestGrid.
\end{acknowledgments}
\bibliography{nanoFCCarxiv.bib}

\end{document}